\documentclass[11pt,a4paper]{article}
\usepackage{amsfonts}
\usepackage[dvips]{graphicx}
\usepackage{epsfig}
\usepackage{color}
\usepackage{amsmath,amssymb}
\usepackage{hyperref}
\newcommand{\arX}[1]{\href{http://arxiv.org/abs/#1}{{\ttfamily\color{blue} arXiv:#1}}}
\usepackage[left=2.27cm,right=2.27cm,top=2.27cm,bottom=2.27cm]{geometry}
\begin{document}

\title{\textbf{De Sitter solutions in Einstein-Gauss-Bonnet gravity}}

\author{Sergey~Vernov\footnote{E-mail address:
svernov@theory.sinp.msu.ru} \ and \ Ekaterina~Pozdeeva\footnote{E-mail address:
pozdeeva@www-hep.sinp.msu.ru}\vspace*{2.72mm} \\
\small  Skobeltsyn Institute of Nuclear Physics, Lomonosov Moscow State University,\\
\small Leninskie Gory 1, Moscow  119991,  Russia}
\date{ \ }

\maketitle

\begin{abstract}
De Sitter solutions play an important role in cosmology because the knowledge of unstable de Sitter solutions can be useful to describe inflation, whereas stable de Sitter solutions are often used in models of late-time acceleration of the Universe. The Einstein--Gauss--Bonnet gravity cosmological models are actively used both as inflationary models and as dark energy models.
To modify the Einstein equations one can add a nonlinear function of the Gauss--Bonnet term or a function of the scalar field multiplied on the Gauss--Bonnet term. The effective potential method essentially simplifies the search and stability analysis of de Sitter solutions, because the stable de Sitter solutions correspond to minima of the effective potential.\\[2.7mm]
\textbf{Keywords:} Einstein--Gauss--Bonnet gravity; \  de Sitter solution; \ stability.
\end{abstract}

\section{Introduction}
 Cosmological models with scalar fields play a central role in the description of the global evolution of the Universe. In particular, modified gravity models with the Ricci scalar multiplied by a function of the scalar field are very popular~\cite{Book-Capozziello-Faraoni,Capozziello:2011et,Fujii_Maeda,Faraoni}. These models are quite natural because quantum corrections to the effective action with a minimally coupled scalar field include nonminimal coupling terms~\cite{ChernikovTagirov,Tagirov,Callan:1970ze}. Many inflationary models that connect cosmology and particle physics include nonminimally coupled scalar fields~\cite{nonmin-quant,higgsinf_0,higgsinf_1,higssinflRG_0,higssinflRG_2,Lerner,higgsinf_2,KaiserHiggs,BezrukovRev,nonmin-infl1,nonmin-infl3,nonmin-infl4,EOPV2014,EOPV2015,PV2017,Dubinin:2017irg,Kamenshchik:2019rot}.

It is well-known that one can add the Gauss--Bonnet term to the Hilbert--Einstein Lagrangian of the General Relativity and it does not change the equations of motion. On the other hand, this term multiplied by some nonconstant function of a scalar field modifies the equations of motion. Models with both the Ricci scalar and the Gauss--Bonnet term multiplied by some functions of the scalar field are natural generalizations of the models with a minimal coupling~\cite{vandeBruck:2015gjd,JoseMathew,Pozdeeva:2021iwc}.
Furthermore, models with a nonlinear function of the Gauss--Bonnet term can be rewritten in the equivalent form that includes a scalar field without kinetic term~\cite{Nojiri:2005vv,Cognola:2006eg,Cognola:2006sp,Nojiri:2017ncd}.

The cosmological models with the Gauss--Bonnet term are motivated by the string theory~\cite{Cognola:2006sp,Nojiri:2017ncd,Antoniadis:1993jc,Kawai1998,Hwang:2005hb,Hwang:2005hb2,ss,Tsujikawa:2006ph} and are actively used  for describing of both the early Universe evolution (inflation)~\cite{vandeBruck:2015gjd,JoseMathew,Pozdeeva:2021iwc,Guo:2010jr,Jiang:2013gza,Koh,DeLaurentis:2015fea,Koh:2016abf,Oikonomou:2017ppp, Wu:2017joj,Nozari:2017rta,Chakraborty:2018scm,Yi:2018gse,Odintsov:2018,Yi:2018dhl,Fomin:2019yls,
Odintsov:2020sqy,Odintsov:2020sqy2,Fomin:2020hfh,Pozdeeva:2020shl,Pozdeeva:2020apf,Oikonomou:2020tct,Pozdeeva:2021nmz} and the current dark energy dominated epoch~\cite{Nojiri:2005vv,Cognola:2006eg,Cognola:2006sp,Nojiri:2017ncd,ss,Tsujikawa:2006ph,Sami:2005zc,Nojiri:2009xh,Elizalde:2010jx,Saez-Gomez,Cruz-Dombriz,Benetti:2018zhv,Navo:2020eqt,Odintsov:2020vjb,Odintsov:2021nim}.
Note that both stages of the Universe evolution are characterized by the quasi de Sitter accelerated expansion of the Universe. So, it is important to have an effective method for the searching of de Sitter solutions and the study of their stability. The method proposed in~\cite{Pozdeeva:2019agu} solves this problem for the Gauss--Bonnet model with the standard scalar field. It is a generalization of the effective potential method for models with scalar field nonminimally coupled to the curvature only~\cite{Skugoreva:2014gka,Pozdeeva:2016cja,Jarv:2021qpp}. We have shown in Ref.~\cite{Pozdeeva:2021iwc} that the effective potential is a useful tool to generalize the known inflationary models with the Gauss--Bonnet~term. On the one hand, the effective potential allows to rule out models with a stable de Sitter solution, for which it is difficult to construct an inflationary scenario with a graceful exit. Models with unstable de Sitter solutions or without an exact de Sitter solution are more suitable. On the other hand, the scalar spectral index $n_s$ and the amplitude of the scalar perturbations $A_s$ as functions of the e-folding number can be expressed via derivatives of the effective potential~\cite{Pozdeeva:2020apf,Pozdeeva:2021iwc,Pozdeeva:2021nmz}.

In this paper, we generalize the effective potential method on models with nonlinear functions of the Gauss--Bonnet term $\mathcal{G}$ that can be presented as models with a scalar field without kinetic term. We also show that the situation is more difficult in the case of a phantom scalar field.

The paper is organized as follows. In Section \ref{sec2}, we remind the evolution equations of the model considered and the standard way for the search of de Sitter solutions.  In \mbox{Section \ref{sec3}}, we analyze the stability of de Sitter solutions due to the effective potential. A way of constructing different models with the same structure of de Sitter solutions is proposed in Section~\ref{sec4}. In Section \ref{sec5}, we analyze the stability of de Sitter solutions in the proposed type of $\mathcal{L}(R,\mathcal{G})$ gravity models. Section~\ref{sec6} is devoted to our conclusions.

\section{Models the Gauss--Bonnet Term}\label{sec2}

Let us consider the  model with the Gauss--Bonnet term described by the following~action:
\begin{equation}
\label{action}
S=\!\int\! d^4 x\sqrt{-g}\left[UR-\frac{c}{2}g^{\mu\nu}\partial_{\mu}\phi\partial_{\nu}\phi- V-F\mathcal{G}\right],
\end{equation}
where the functions $U(\phi)$, $V(\phi)$, and $F(\phi)$ are double differentiable ones, $c$ is a constant, $R$ is the Ricci scalar and $\mathcal{G}$ is the Gauss--Bonnet term,
\begin{equation*}
\mathcal{G}=R^2-4R_{\mu\nu}R^{\mu\nu}+R_{\mu\nu\alpha\beta}R^{\mu\nu\alpha\beta}.
\end{equation*}

As known~\cite{Nojiri:2017ncd,Cognola:2006eg}, the action
\begin{equation}
\label{actionl}
S=\!\int\! d^4 x\sqrt{-g}\left[U_0R+\mathcal{L}(\mathcal{G})\right],
\end{equation}
where $U_0$ is a constant and $\mathcal{L}(\mathcal{G})$ is a double differentiable function, can be rewritten in the form of action (\ref{action}) with $c=0$:
 \begin{equation}
\label{action2}
S=\!\int\! d^4 x\sqrt{-g}\left[U_0R+\mathcal{L}'(\phi)(\mathcal{G}-\phi)+\mathcal{L}(\phi)\right],
\end{equation}
where a prime denotes the derivatives with respect to $\phi$. Varying action (\ref{action2}) over $\phi$, one gets
$\phi=\mathcal{G}$ and the initial $\mathcal{L}(\mathcal{G})$ model.

In the spatially flat Friedmann--Lema\^{i}tre--Robert\-son--Walker metric with the interval
\begin{equation}
\label{metric}
ds^2={}-dt^2+a^2(t)\left(dx_1^2+dx_2^2+dx_3^2\right),
\end{equation}
where $a(t)$ is the scale factor,
one gets the following evolution equations~\cite{Oikonomou:2020tct}:
\begin{equation}\label{Equ00}
    6H^2U+6HU'\dot{\phi}=\frac{c}{2}{\dot{\phi}}^2+V+24H^3F'\dot{\phi},
\end{equation}
\begin{equation}\label{EquH}
    4\left(U-4H\dot{F}\right)\dot{H}={}-c\dot{\phi}^2-2\ddot{U}+2H\dot{U}+8H^2\left(\ddot{F}-H\dot{F}\right),
   \end{equation}
\begin{equation}
\label{equphi}
c\ddot\phi+3cH\dot\phi-6\left(\dot H+2H^2\right)U'+V'+24H^2F'\left(\dot H +H^2\right)=0,
\end{equation}
where  $H=\dot{a}/a$ is the Hubble parameter, dots and primes denote the derivatives with respect to the cosmic time and  the scalar field $\phi$, respectively.
At $c=1$, these equations have been investigated in many papers (see, for example~\cite{vandeBruck:2015gjd,Pozdeeva:2019agu,Oikonomou:2020tct}).

To find de Sitter solutions with a constant~$\phi$ in the model (\ref{actionl}) we substitute $\phi=\phi_{dS}$ and $H=H_{dS}$ into Equations~(\ref{Equ00}) and (\ref{equphi}).
A de Sitter solution does not depend on the value of $c$, so we obtain the same results as in the case of $c=1$ considered in~\cite{Pozdeeva:2019agu}:
\begin{equation}
\label{HdS}
H_{dS}^2=\frac{V_{dS}}{6U_{dS}}
\end{equation}
and
\begin{equation}
\label{equdS}
F'_{dS} = \frac{3U_{dS}(2U'_{dS}V_{dS}-V'_{dS}U_{dS})}{2V_{dS}^2},
\end{equation}
where $A_{dS}\equiv A(\phi_{dS})$ for any function $A$.
Therefore, for arbitrary functions $U(\phi)$ and $V(\phi)$ with  $V_{dS}U_{dS}>0$, we can choose
$F(\phi)$ such that the corresponding point becomes a de Sitter solution, with the Hubble parameter defined by Equation~(\ref{HdS}). We always choose that $H_{dS}>0$.

\section{Stability of de Sitter Solutions}\label{sec3}

To analyze the stability of a de Sitter solution we transform Equations~(\ref{EquH}) and (\ref{equphi}) to the following dynamical system:
\begin{equation}
\label{DynSYS}
\begin{split}
\dot\phi=&\psi,\\
\dot\psi=&\frac{1}{2\left(\tilde{B}-4cF'H\psi\right)}\left\{2H\left[3B+4F'V'-6{U'}^2-6cU\right]\psi- 2\frac{V^2}{U}X
\right.\\
+&\left. \left[12H^2\left[\left(2U''+3c\right)F'+2U'F''\right]
- 96F'F''H^4-3\left(2U''+c\right)U'\right]\psi^2 \right\},\\
\dot H=&\frac{1}{4\left(\tilde{B}-4cF'H\psi\right)}\left\{8c\left(U'-4F'H^2\right)H\psi\right.\\
-&\left. 2\frac{V^2}{U^2}\left(4F^\prime H^2-U'\right) X+ \left(8F''H^2-2{U''}-c\right)c\psi^2\right\},
\end{split}
\end{equation}
where
\begin{equation}
\label{B}
\tilde{B}=3\left(4H^2F^\prime -U^\prime \right)^2+cU,
\end{equation}
\begin{equation}
\label{X}
X=\frac{U^2}{V^2}\left[24 H^4 F^\prime -12 H^2 U^\prime +V^\prime \right].
\end{equation}

In the case $c=0$, the last equation is essentially simplified:
\begin{equation}
\label{equHc0}
\dot H=\frac{24H^4F'-12H^2U'+V'}{6\left(U'-4H^2F'\right)}.
\end{equation}

At a de Sitter point system~(\ref{DynSYS}) is
\begin{equation*}
\dot\phi=0,\quad\dot\psi=0, \quad \dot H=0,
\end{equation*}
that corresponds to $X_{dS}=0$.

In Ref.~\cite{Pozdeeva:2019agu}, the effective potential has been proposed  for models with the Gauss--Bonnet term:
\begin{equation}
\label{V_eff}
    V_{eff}={}-\frac{U^2}{V}+\frac{2}{3}F.
\end{equation}

Using Equations~(\ref{HdS}) and (\ref{X}), we obtain
\begin{equation}
\label{XdS}
    X_{dS}=\frac{2}{3}F^\prime_{dS}-2 \frac{U^\prime_{dS}U_{dS}}{V_{dS}}+\frac{V^\prime_{dS} U_{dS}^2}{V^2_{dS}}=V'_{eff}(\phi_{dS})=0,
\end{equation}
therefore, de Sitter solutions correspond to extremum points of the effective potential~$V_{eff}$.

To investigate the Lyapunov stability of a de Sitter solution we use the following expansions:
\begin{equation}
\label{H1phi1}
H(t)=H_{dS}+\varepsilon H_1(t),\qquad \phi(t)=\phi_{dS}+\varepsilon \phi_1(t),\qquad \psi(t)=\varepsilon \psi_1(t),
\end{equation}
where $\varepsilon$ is a small parameter. Therefore,
\begin{equation}
\label{X1}
X=\varepsilon (X_{,H}H_1+X_{,\phi}\phi_1)+{\mathcal O}(\varepsilon^2),
\end{equation}
 where
\begin{equation*}
X_{,H}=\left.\frac{\partial X}{\partial H}\right|_{\phi=\phi_{dS}}=\frac{4\sqrt{6}}{V_{dS}^{5/2}}U_{dS}^{3/2}\left(U'_{dS}V_{dS}-V_{dS}'U_{dS}\right),
\end{equation*}
\begin{equation*}
X_{,\phi}=\left.\frac{\partial X}{\partial \phi}\right|_{\phi=\phi_{dS}}=\frac{1}{V_{dS}^2}\left(\frac{2}{3}F_{dS}''V_{dS}^2-2U_{dS}''U_{dS}V_{dS}+V_{dS}''U_{dS}^2\right).
\end{equation*}

The functions $H_1(t)$, $\phi_1(t)$, and $\psi_1(t)$ are connected by Equation~(\ref{Equ00}):
\begin{equation}
\label{H1}
    H_1(t)=\frac{V'_{dS}U_{dS}-U'_{dS}V_{dS}}{2U_{dS}V_{dS}}\left(H_{dS}\phi_1(t)-\psi_1(t)\right).
\end{equation}
This expression does not depend on the value of $c$ and coincides with the corresponding expression obtained in Ref.~\cite{Pozdeeva:2019agu}.

Substituting (\ref{H1phi1})--(\ref{H1}) into Equation~(\ref{DynSYS}) in the first order of $\varepsilon$,
we obtain the following system of two linear differential equations:

\begin{equation}
\label{equphi1}
\dot\phi_1=\psi_1,
\end{equation}

\begin{equation}
\label{equpsi1}
\dot\psi_1={}-\frac{\left[2F_{dS}''V_{dS}^3-6U_{dS}''U_{dS}V_{dS}^2+3V_{dS}''U_{dS}^2V_{dS}
-6\left(U'_{dS}V_{dS}-V'_{dS}U_{dS}\right)^2\right]}{3U_{dS}V_{dS}B_{dS}}\phi_1
-\frac{\sqrt{6U_{dS}V_{dS}}}{2U_{dS}}\psi_1,
\end{equation}
where
\begin{equation}\label{BdS}
    B_{dS}=\frac{3}{V_{dS}^2}\left(V'_{dS}U_{dS}-U'_{dS}V_{dS}\right)^2+cU_{dS}.
\end{equation}

This system can be rewritten in the matrix form:
\begin{equation}\label{syst1}
    \left( \begin{array}{c} \dot\phi_1 \\ \dot\psi_1 \end{array} \right)=
     \left( \begin{array}{c} {A}_{11} \qquad {A}_{21}  \\ {A}_{12} \qquad {A}_{22} \end{array}\right)
     \left( \begin{array}{c} \phi_1 \\ \psi_1 \end{array} \right)
\end{equation}
where the matrix
\begin{equation*}
A=\begin{array}{||cc||}0,&1\\{}
{}-\frac{V_{dS}^2V_{eff}''(\phi_{dS})}{U_{dS}B_{dS}},\quad &
-3H_{dS}\,
\end {array}
\end{equation*}

The general solution of system~(\ref{syst1}) has the following form:
\begin{equation}\label{solphi1}
  \phi_1=c_{11}\mathrm{e}^{-\lambda_{-}t} +c_{21}\mathrm{e}^{-\lambda_{+}t},
\end{equation}
\begin{equation}
\label{solpsi1}
  \psi_1=c_{21}\mathrm{e}^{-\lambda_{-}t} +c_{22}\mathrm{e}^{-\lambda_{+}t},
\end{equation}
where $c_{ij}$ are some constants. Solving the characteristic equation:
\begin{equation}
\det(\tilde{A}-\lambda\cdot I)=\lambda^2-3H_{dS}\lambda+\frac{V_{dS}^2V_{eff}''(\phi_{dS})}{U_{dS}B_{dS}}=0,
\end{equation}
we get the following roots:
\begin{equation}\label{lambda12}
   \lambda_\pm={}-\frac{3}{2}H_{dS}\pm\sqrt{\frac{9}{4}H_{dS}^2-\frac{V_{dS}^2}{U_{dS}B_{dS}}V''_{eff}(\phi_{dS})}\,.
\end{equation}

A de Sitter solution is stable if real parts of both $\lambda_-$ and $\lambda_+$ are negative.
We consider the case
\begin{equation*}
H_{dS}=\sqrt{\frac{V_{dS}}{6U_{dS}}}>0,
\end{equation*}
 hence, $\Re e (\lambda_-)<0$.

In the case of a positive $U_{dS}$, we see that $B_{dS}>0$ for $c \geqslant 0$ and
the condition $\Re e (\lambda_+)<0$ is equivalent to $V''_{eff}(\phi_{dS})>0$.  In the cases $c>0$ and $c=0$, a de Sitter solution is stable if $V''_{eff}(\phi_{dS})>0$ and unstable if $V''_{eff}(\phi_{dS})<0$.

In the case of $c<0$, we see that  $B_{dS}$ can be negative. So, in this case de Sitter solution is stable if the $V''_{eff}(\phi_{dS})B_{dS}>0$.
So, the main result of Ref.~\cite{Pozdeeva:2019agu} can be generalized on the case $c=0$ without any correction, whereas  the condition should be change to  $V''_{eff}(\phi_{dS})B_{dS}>0$
in the case of $c<0$ that corresponds to a phantom scalar field $\phi$.

\section{Different Models with the Same Structure of de Sitter Solutions}\label{sec4}

It is evident that any change of functions $U$ and $V$ can be compensated by the change of function $F$ such that the first derivative of the effective potential does not change. This property can be used for the generalization of the inflationary scenarios~\cite{Pozdeeva:2021iwc}. Let us consider a more nontrivial question: how one can change the functions $U$ and $V$ only to obtain the model with the same structure of de Sitter solutions. To be concrete we seek for models with the same values of $\phi_{dS}$ and the same stability properties of de Sitter solutions, but maybe with different values of $H_{dS}$.

 Considering the stability analysis of de Sitter solutions, we define the effective potential as such a function that its minima correspond to the stable de Sitter solutions and maxima correspond to unstable de Sitter solutions. For this reason, the effective potential is not unique. We can add a constant to it or multiply it on a positive number.
If we transform functions $U(\phi)$ and $V(\phi)$ to functions $\tilde{U}(\phi)=f(\phi)U(\phi)$ and
\begin{equation}
\tilde{V}(\phi)=\frac{V(\phi)f(\phi)^2U(\phi)^2}{W(\phi)V(\phi)+U(\phi)^2},
\end{equation}
then the effective potentials of the original and transformed models are connected as follows:
\begin{equation}
\tilde{V}_{eff}=V_{eff}+W.
\end{equation}

Therefore, the structure of de Sitter solutions does not change if the function\linebreak  $W=CV_{eff}+W_0$, where a constant $C>-1$ and $W_0$ is an arbitrary constant. Furthermore, one should check that  $\tilde{V}(\phi_{dS})>0$ and  {$\tilde{U}(\phi_{dS})>0$} for all de Sitter solutions.

Moreover, if $V_{eff}(\phi)>0$ for any $\phi$, then the functions $V_{eff}(\phi)^n$ and $-1/V_{eff}(\phi)^n$, where $n$ is a natural number, can be considered as new effective potentials.
Different transformations of the effective potential can be combined, for example, functions $-1/(V_{eff}+W)$ and $(V_{eff}+W)^2$ can be considered as effective potentials if $V_{eff}>-W$ for any $\phi$. So, transformations of the model with $\tilde{V}_{eff}=(V_{eff}+W)^2$ or $\tilde{V}_{eff}=-1/(V_{eff}+W)$ does not change the structure of de Sitter solutions.

\section{Examples of $\boldsymbol{\mathcal{L}(R,\mathcal{G})}$ Models}\label{sec5}
\subsection{Evolution Equations}
Several examples of models with an ordinary scalar field coupled with the Gauss--Bonnet term have been considered in Refs.~\cite{Pozdeeva:2020apf,Pozdeeva:2019agu}. In this paper, we consider examples
in the case of $c=0$, namely, the case of $\mathcal{L}(R,\mathcal{G})$ model, described by the following action:
\begin{equation}
\label{actionFRG}
S=\!\int\! d^4 x\sqrt{-g}\left[U_0R+\mathcal{L}(C_1R+C_2\mathcal{G})\right],
\end{equation}
where $\mathcal{L}$ is a double differentiable function, $U_0$, $C_1$, and $C_2$ are constants. A linear function $\mathcal{L}$ corresponds to the General Relativity, whereas a nonlinear function $\mathcal{L}$ corresponds to the modified gravity. Note that $\mathcal{L}(R)$ and $\mathcal{L}(\mathcal{G})$ models are particular cases of the model~considered.

In the case of a nonlinear function $\mathcal{L}$, action (\ref{actionFRG}) can be rewritten in the following~form:
 \begin{equation}
\label{action3}
S=\!\int\! d^4 x\sqrt{-g}\left[U_0R+\mathcal{L}'(\phi)(C_1R+C_2\mathcal{G}-\phi)+\mathcal{L}(\phi)\right].
\end{equation}

Varying action (\ref{action3}) over $\phi$, one gets
$\phi=C_1R+C_2\mathcal{G}$ and the initial $\mathcal{L}(R,\mathcal{G})$ model with action~(\ref{actionFRG}).
Action (\ref{action3}) is a particular case of action (\ref{action}) with the functions:
\begin{equation}\label{UFVc0}
    U=U_0+C_1{\mathcal{L}'},\quad V=\phi\mathcal{L}'-\mathcal{L}, \quad F=C_2\mathcal{L}', \quad c=0.
\end{equation}

The effective  {potential} 
is the following combination of the function $\mathcal{L}$ and its first derivative:
\begin{equation}
\label{VeffFRG}
    V_{eff}=\frac{2}{3}C_2\mathcal{L}'-\frac{\left(U_0+C_1\mathcal{L}'\right)^2}{\phi\mathcal{L}'-\mathcal{L}}.
\end{equation}
So,
\begin{equation}
\label{dVeffL}
V_{eff}'=\frac{\mathcal{L}''}{\left({\phi\mathcal{L}'-\mathcal{L}}\right)^2}\left[(3C_1^2-2C_2\phi)\left(\phi{\mathcal{L}'}-2{\mathcal{L}}\right){\mathcal{L}}' -2C_2{\mathcal{L}'}^2-6C_1U_0{\mathcal{L}}-3U_0^2\phi\right]
\end{equation}
and a point with $\mathcal{L}''=0$ correspond to de Sitter solutions if $\phi\mathcal{L}'>\mathcal{L}$ and $U_0+C_1{\mathcal{L}'}>0$. Furthermore, the second multiplied can be equal to zero that also can correspond to a de Sitter solution.
We explorer de Sitter solutions in detail in a few examples of $\mathcal{L}(R,\mathcal{G})$ models.

\subsection{The Function $F$ in a Role of the Effective Potential}

If $V=CU^2$, where $C$ is a constant, then $V'_{eff}=2F'/3=2C_2\mathcal{L}''/3$ and the function $\mathcal{L}'$ plays the role of the effective potential.
For the considering $\mathcal{L}(R,\mathcal{G})$ models the condition $V=CU^2$ is the following first order differential equation
\begin{equation}\label{equL}
    CC_1^2 {\mathcal{L}'}^2+(2CC_1U_0-\phi){\mathcal{L}'}+CU_0^2+\mathcal{L}=0.
\end{equation}

This equation has two solutions:
\begin{equation}
\mathcal{L}_1=A\phi-C(C_1A+U_0)^2,\qquad \mathcal{L}_2=\frac{1}{4CC_1^2}\phi^2-\frac{U_0}{C_1}\phi,
\end{equation}
where $A$ is an integration constant. The function $\mathcal{L}_1$ is a linear one, so this case is the General Relativity model with the cosmological constant. In the case of the function $\mathcal{L}_2$, the function $F$ is a linear one and this model has no de Sitter solution.

\subsection{The Case of a Power Function $\mathcal{L}$ of the Gauss--Bonnet Term}

Let $\mathcal{L}=C\mathcal{G}^\alpha$, where $C$ and $\alpha$ are constants.  Substituting this function with $C_1=0$ and $C_2=1$ into Equation~(\ref{dVeffL}), we get
\begin{equation}
V_{eff}'=\frac{2\alpha\left(2C^2(\alpha-1)^2\phi^{2\alpha}+3U_0^2\phi\right)}{6(C(\alpha-1)\phi^{\alpha+2})},
\end{equation}

There is no de Sitter solutions at $\phi_{dS}>0$. So, we consider only such values of $\alpha$ that $\phi^\alpha$ is real and negative at $\phi<0$ and obtain
\begin{equation*}
\phi_{dS}={}-\left(\frac{3U_0^2}{2C^2(\alpha-1)^2}\right)^{\frac{1}{2\alpha-1}}.
\end{equation*}

  Let us consider several examples:
\begin{enumerate}
  \item At $\alpha=2$, we get de Sitter solution
  \begin{equation}
  \phi_{dS}=-\left(\frac{3U_0^2}{2C^2}\right)^{1/3}
  \end{equation}
  \end{enumerate}
  and the potential $V=C\phi^2$, so, $C>0$ is a necessary condition for de Sitter solutions existence.

  The effective potential
\begin{equation}
  V_{eff}=\frac{4C\phi}{3}-\frac{U_0}{C\phi^2},
 \end{equation}
 and its second derivative is negative: $V^{\prime\prime}_{eff}={}-\frac{6U_0}{C\phi^4}$. Therefore, we can conclude that the de Sitter solution is unique and unstable.
\begin{enumerate}
  \item[2.] At $\alpha=3$ we get
\begin{equation}
\phi_{dS}={}-\left(\frac{3U_0^2}{8C^2}\right)^{1/5}.
\end{equation}
  \end{enumerate}

The potential
\begin{equation}
V=2C\phi^3.
\end{equation}

The condition $V(\phi_{dS})>0$ demands $C<0$ for $\phi_{dS}>0$. Using
\begin{equation}
V^{\prime\prime}_{eff}(\phi_{dS})=20C,
\end{equation}
we get that the considered model with $\alpha=3$ and  $C<0$ has an unstable de Sitter solution.
\begin{enumerate}
  \item[3.] Let us consider the case of $\alpha=1/3$. Similar models have been proposed in~\cite{Nojiri:2017ncd}.
For
\begin{equation}
\phi_{dS} ={}-\frac{512C^6}{19683U_0^6},
\end{equation}
\end{enumerate}
we obtain
\begin{equation}
V(\phi_{dS})=\frac{16C^3}{81U_0^2}\quad\mbox{and} \quad V''_{eff}(\phi_{dS})={}-\frac{3486784401U_0^{16}}{4194304C^{15}}.
\end{equation}

So, a de Sitter solution exist if and only if $C>0$ and it is unstable.

\subsection{The Case of a Quadratic Polynomial $\mathcal{L}$}
\subsubsection{Equation for $\phi_{dS}$}
Let us consider the case of
\begin{equation}
\label{quadratic}
   \mathcal{L}=b_2\phi^2+b_1\phi+b_0,
\end{equation}
where $b_i$ are constants, $b_2\neq 0$. In this case,
\begin{equation}\label{UVF2}
    U=2C_1b_2\phi+Q_0,\quad  V=b_2\phi^2-b_0, \quad F=C_2(2b_2\phi+b_1),
\end{equation}
where $Q_0=U_0+b_1C_1$. Note that the function $F$ is defined up to a constant, so we can put $b_1=0$ without loss of generality.

De Sitter solution corresponds to $\phi_{dS}$ that is a solution of the following equation:
\begin{equation}
\label{EquphidS}
2C_2b_2^2\phi_{dS}^4+2(3C_1b_2Q_0-2C_2b_0b_2)\phi_{dS}^2+3(4C_1^2b_0b_2+Q_0^2)\phi_{dS}+6C_1b_0Q_0+2C_2b_0^2=0.
\end{equation}

Let us consider a few interesting particular cases of this equation.

\subsubsection{The Case $\mathcal{L}(R)$ Gravity}
If $C_2=0$, then the model has no Gauss--Bonnet term and is a $\mathcal{L}(R)$ gravity model with the effective potential
\begin{equation}
V_{eff}=\frac{\left(2C_1b_2\phi+\tilde{U}_0\right)^2}{b_0-b_2\phi^2}\,.
\end{equation}

Equation (\ref{EquphidS}) is a quadratic one and has the following two solutions:
\begin{equation}
\phi_1 = {}-\frac{2C_1b_0}{\tilde{U}_0},\qquad \phi_2 = {}-\frac{\tilde{U}_0}{2C_1b_2}\,.
\end{equation}

The point $\phi=\phi_2$ does not correspond to de Sitter solution because of $U(\phi_2)=0$. At $\phi=\phi_1$, we get
\begin{equation}
\label{D2Veff}
\left.V_{eff}''\right|_{\phi=\phi_1}={}-\frac{2b_2\tilde{U}_0^4}{b_0^2\left(4C_1^2b_0b_2-\tilde{U}_0^2\right)},
\end{equation}

At de Sitter point $\phi=\phi_1$, we obtain
\begin{equation}
U_{dS}=\frac{\tilde{U}_0^2-4C_1^2b_0b_2}{\tilde{U}_0},\qquad V_{dS}=\frac{4b_2C_1^2b_0^2}{\tilde{U}_0^2}-b_0.
\end{equation}

From Equation~(\ref{HdS}), we get
\begin{equation}\label{H2FR}
    H^2_{dS}={}-\frac{b_0}{6\tilde{U}_0},
\end{equation}
a de Sitter solution exists only if $b_0\tilde{U}_0<0$. Note that the Starobinsky $R^2$ inflationary model~\cite{Starobinsky:1980te,Starobinsky:1982,Starobinsky:1983} does not include the cosmological constant, so, $b_0=0$. In this case, a de Sitter solution does not exist. The potential $V>0$ for all values of $\phi$ if $b_2>0$ and $b_0<0$. In this case, the de Sitter solution is stable.

\subsubsection{The Case $\mathcal{L}(\mathcal{G})$ Gravity}
 If $C_1=0$, then the model is a $\mathcal{L}(\mathcal{G})$ gravity models with the effective potential
\begin{equation}
\label{VeffG2}
    V_{eff}'=\frac{2b_2\left(2C_2b_2^2\phi^4-4C_2b_0b_2\phi^2+3U_0^2\phi+2C_2b_0^2\right)}{3(b_2\phi^2-b_0)^2}
\end{equation}
The value of $\phi$ at a de Sitter point is a real solution of the following equation:
\begin{equation}
\label{equdSL}
2C_2\left(b_0-b_2\phi_{dS}^2\right)^2+3U_0^2\phi_{dS}=0.
\end{equation}

In the case of $b_0=0$, one gets the following real solutions of Equation~(\ref{equdSL}):
\begin{equation}
\phi_1=0,\qquad \phi_2={}-\frac{\left(12U_0^2C_2^2b_2\right)^{1/3}}{2C_2b_2}.
\end{equation}

Note that $\phi_1$ does not correspond to a de Sitter solution, because $V(\phi_1)=0$. So, we get the unique de Sitter solution $\phi_{dS}=\phi_2$.

At the de Sitter point, the second derivative of the effective potential is
\begin{equation}
V''_{dS}={}-\frac{2}{3}\left(144b_2^5C_2^4U_0^2\right)^{1/3}.
\end{equation}

Furthermore, we demand that $V_{dS}=b_2\phi_{ds}^2>0$, so, $b_2>0$ and we get a model with one unstable de Sitter solution.

\subsubsection{The Case of the Absence of the Cosmological Constant}
In the case of $b_0=0$, the potential $V=b_2\phi^2$, therefore, $\phi_{dS}\neq 0$ and $b_2>0$ are necessary condition for de Sitter solution existence.
Assuming that $Q_0\neq 0$, we obtain the first derivative of the effective potential in the following form:
\begin{equation}
\label{DVeffb0}
    V'_{eff}=\frac{2Q_0^2\left(2B_1\phi^3+6B_2\phi+3\right)}{3b_2\phi^3},
\end{equation}
where
\begin{equation}
  \label{CB}
    B_1\equiv\frac{b_2}{Q_0}C_1,\qquad B_2=\frac{b_2^2}{Q_0^2}C_2\,.
\end{equation}

The effective potential can be multiplied on any positive constant, so we can consider the function
\begin{equation}
J(\phi)= \frac{2B_1\phi^3+3(2B_2\phi+1)}{\phi^3}
\end{equation}
instead of $V'_{eff}$ and the number of its zeros depends on values of parameters $B_1$ and $B_2$ only.

The function
\begin{equation}
\label{UB}
U= Q_0(2B_2\phi+1)
\end{equation}
is not always positive, so it is possible that an extremum of the effective potential does not corresponds to de Sitter solution.
The sign of $U(\phi_{dS})$  {depends on} the sign of the parameter $Q_0$, whereas $V'_{eff}$ does not depend on the sign of the parameter $Q_0$.
Note that the functions $U$ and $J$ are not equal to zero at the same point, because $B_1\neq 0$.

\begin{figure}[h]
\includegraphics[width= 7cm, height= 8.1cm]{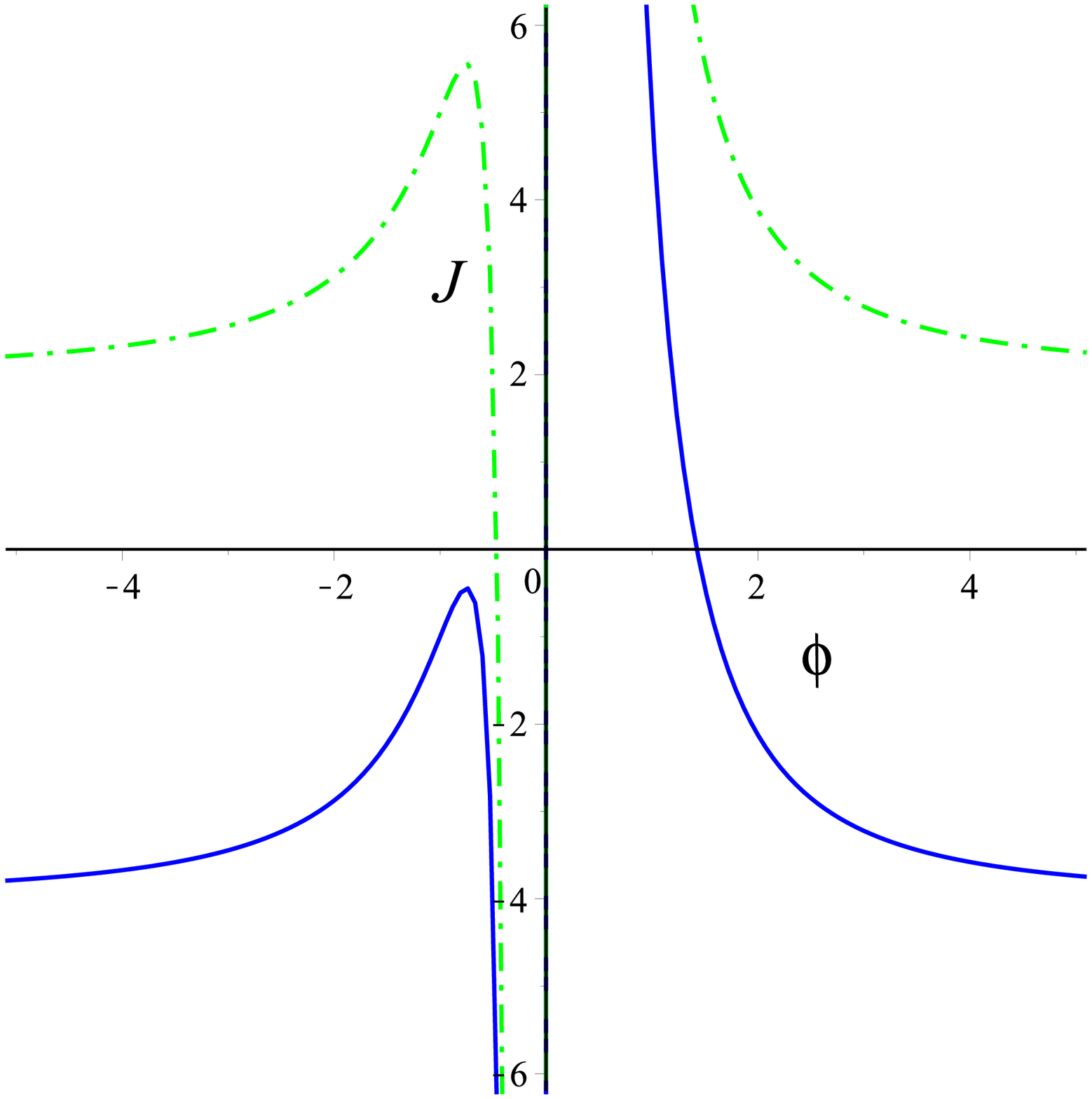} \ \ \ $\qquad$ \ \
\includegraphics[width= 7cm, height= 8.1cm]{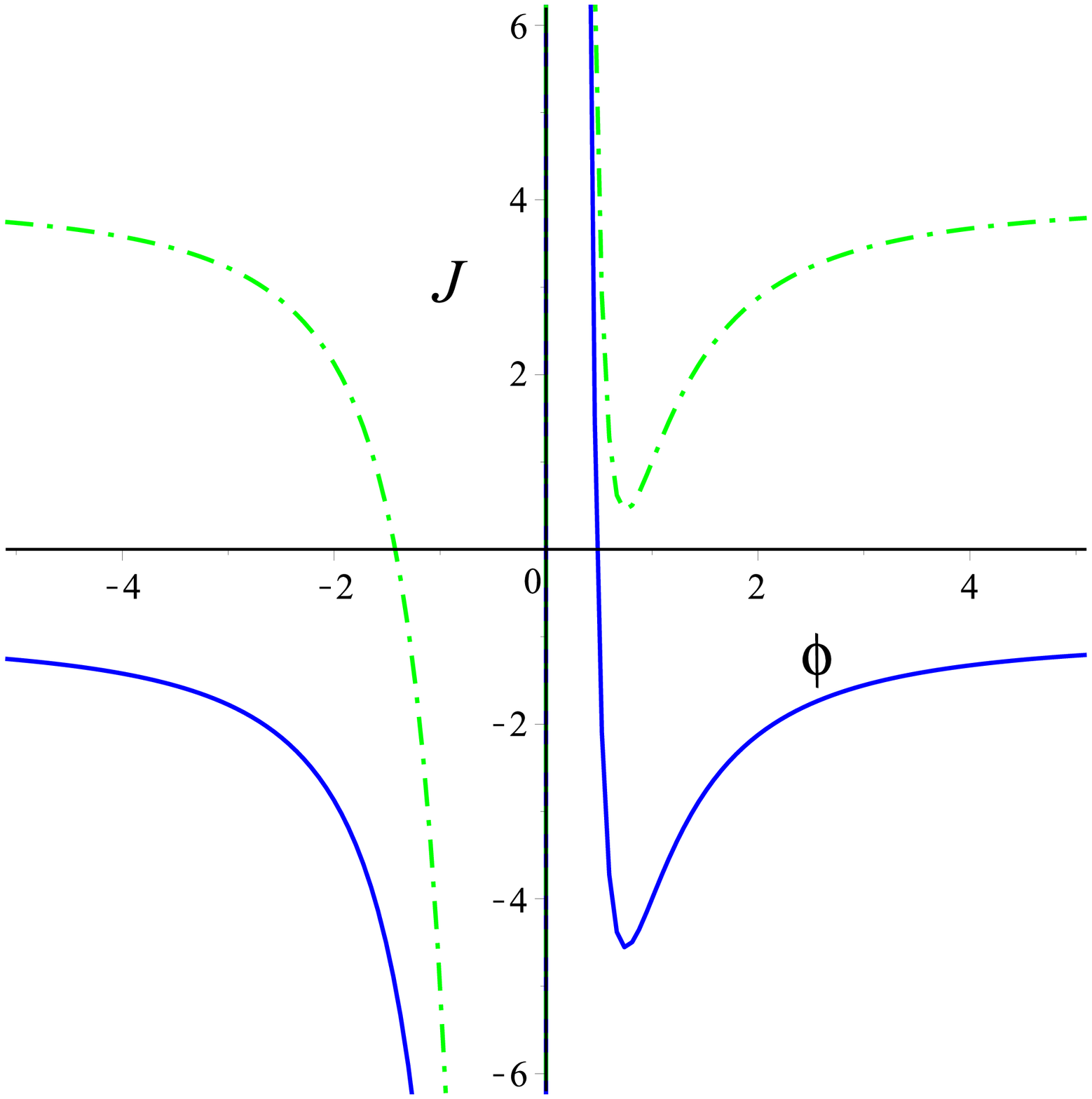}
\caption{ {The function} $J(\phi)$ for different values of parameters: $B_2=1$, $B_1=-2$ (blue solid curve) and $B_1=1$ (green dash-dot curve) (\textbf{left});
$B_2=-1$, $B_1=-1/2$ (blue solid curve) and $B_1=2$ (green {dash-dot curve) (\textbf{right}).}
\label{Jphi}}
\end{figure}

The same values of parameters $B_1$ and $B_2$ correspond both to models with one de Sitter solution and to models without de Sitter solution in dependence of the sign of $Q_0$.
A decreasing behavior of the function $J$ in the neighborhood of $J=0$ corresponds to an unstable de Sitter solution. In Figure~\ref{Jphi}, blue solid curves and green dash-dot curves correspond to models either with one unstable de Sitter solution or without de Sitter solutions in dependence of the sign of $U(\phi_{dS})$.

\begin{figure}[h]
\includegraphics[width= 7cm, height= 8.1cm]{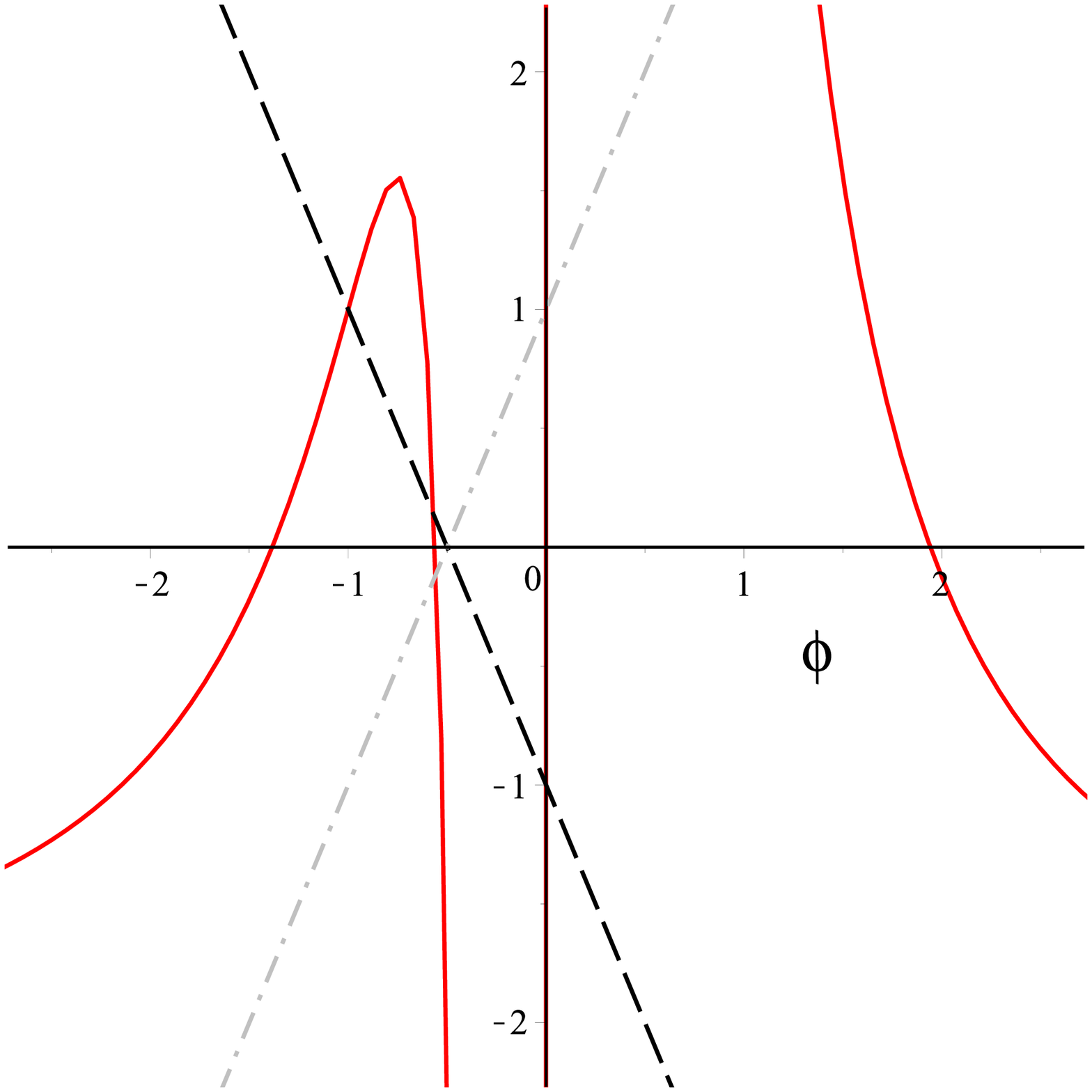} \ \ \ $\qquad$ \ \
\includegraphics[width= 7cm, height= 8.1cm]{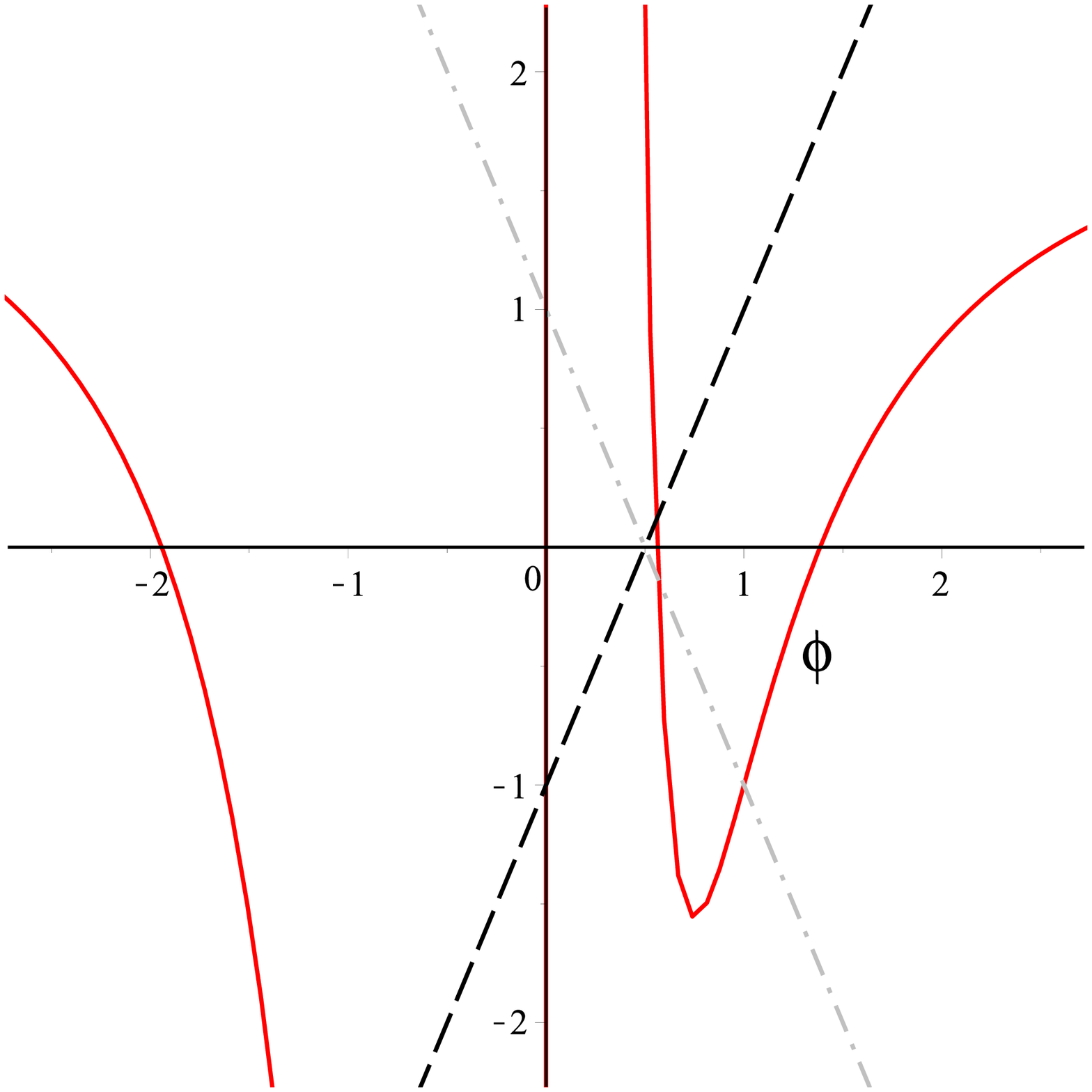}
\caption{ {The functions} $J(\phi)$ and $U(\phi)$ for $B_2=1$ (\textbf{left}) and $B_1=-1$ (\textbf{right}). Red solid curves show the function $J(\phi)$.
Black dash curves show the function $U(\phi)$ at $Q_0=-1$ and grey dash-dot curves show the function $U(\phi)$ at $Q_0=1$.
\label{Jphi2}}
\end{figure}

In Figure~\ref{Jphi2}, we present the function $J(\phi)$ that has  {three roots} (red solid curves) and the function $U(\phi)$ for the same values of parameters $B_1$ and $B_2$ and the parameter $Q_0=\pm 1$. One can see that the model has one unstable de Sitter solution or one stable and one unstable de Sitter solutions in dependence on the sign of $Q_0$. The crosses of gray and black lines with axis $\phi$ correspond to switch of gravity and antigravity regimes in the model~considered.

\vspace{-6PT}

\section{Conclusions}\label{sec6}
In this paper, we consider de Sitter solutions in models with the Gauss--Bonnet term, including $\mathcal{L}(R,\mathcal{G})$ gravity models.
We investigate evolution equations of a scalar field nonminimally coupled both with the curvature and with the Gauss--Bonnet term and look for the fixed points of scalar field dynamics which correspond to de Sitter solutions. We show that, in the case of a positive coupling function $U(\phi)$, it is possible to introduce an effective potential $V_{eff}$ which can be expressed
through the function $U$ of nonminimal coupling with the curvature, the scalar field potential $V$, and the coupling function with the Gauss--Bonnet term denoted by~$F$. We show that it is convenient to investigate the structure of fixed points using the effective potential because the stable de Sitter solutions correspond to minima of the effective potential.
The existence and stability of de Sitter solutions in the system under consideration can be studied with the help of function $V_{eff}$ since one can get the structure and stability properties of de Sitter solutions using a graphical representation of the effective potential only. It should be noted that the effective potential is not uniquely defined. We can multiply it on a positive constant or add a constant to it. If the effective potential is a positive definite function, then we can consider $V_{eff}^n$ and $-1/V_{eff}^n$, where $n$ is a natural number, like other forms of the effective potential.

In this paper, we show that the effective potential proposed~\cite{Pozdeeva:2019agu} for models with the Gauss--Bonnet term multiplied on a function of the scalar field can be used in $\mathcal{L}(R,\mathcal{G})$ models as well.
To find de Sitter solutions in some $\mathcal{L}(R,\mathcal{G})$ model, we rewrite the action of this model in the form~(\ref{action3}) and construct the corresponding effective potential $V_{eff}$.
A stable de Sitter solutions corresponds $V''_{eff}(\phi_{dS})>0$, where the values of the scalar field at de Sitter point $\phi_{dS}$ is determined by the condition $V'_{eff}(\phi_{dS})=0$.
We have found de Sitter solutions in a few $\mathcal{L}(R,\mathcal{G})$ models to demonstrate the effective potential method.

Note that the proposed effective potential is a  useful tool for the construction of inflationary scenarios in the models with the Gauss--Bonnet term multiplied to a function of the scalar field~\cite{Pozdeeva:2020apf,Pozdeeva:2021iwc,Pozdeeva:2021nmz}. It is interesting that in the slow-roll approximation the scalar spectral index $n_s$ and the amplitude of the scalar perturbations $A_s$ as functions of the e-folding number can be expressed via derivatives of the effective potential given in the form (\ref{V_eff}).
Furthermore, the knowledge of unstable de Sitter solutions can be useful to describe inflation (see, for example, Ref.~\cite{EOPV2014}).
We plan to generalize this approach to inflationary scenarios in $\mathcal{L}(R,\mathcal{G})$ models.

The search for stable de Sitter solutions is important for  {$f(R, \mathcal{G})$} models that explain the late-time
accelerated expansion of the  {Universe} {\cite{Cognola:2006eg,Nojiri:2009xh,Elizalde:2010jx,Cruz-Dombriz,Odintsov:2020vjb,Odintsov:2021nim}}.
 A generic  {$f(R, \mathcal{G})$} action can be transformed
into one linear in  {$R$ and $\mathcal{G}$} by including of two scalar fields, whereas the proposed special type of such
models describing by action (29) can be linearized in  {$R$ and $\mathcal{G}$} by including of one scalar field without
kinetic term. It allows to use the effective potential method and to simplify analysis of the stability
of de Sitter solutions in distinguish to the traditional  {approach} \cite{Cruz-Dombriz}. The proposed  {$\mathcal{L}(R, \mathcal{G})$} models
include not only $F(R)$ and  {$F(\mathcal{G})$} gravity models, but also more complicated models with the  {$\mathcal{L}(R + \mathcal{G})$} function in the action.
We plan to investigate a possibility to describe the late-time accelerated expansion of the Universe
in such types of models taking into account the observation restrictions.

{ \ }

\textbf{{Acknowledgements}}

This work is partially supported by  the Russian Foundation for Basic Research grant No. 20-02-00411.

\end{document}